\documentclass[prl,twocolumn,amsmath,amssymb]{revtex4}
\usepackage{graphicx}

\begin{document}
\title{Entanglement spectrum of one-dimensional extended Bose-Hubbard models}
\author{Xiaolong Deng and Luis Santos}
\affiliation{Institut f\"ur Theoretische Physik, Leibniz Universit\"at Hannover, Appelstr. 2, 30167 Hannover, Germany}

\begin{abstract}

The entanglement spectrum provides crucial information about correlated quantum systems.
We show that the study of the block-like nature of the reduced density matrix in number sectors  
and the partition dependence of the spectrum in finite systems leads to interesting 
unexpected insights, which we illustrate for the 
case of a 1D extended Hubbard model. We show that block symmetry provides an intuitive understanding of 
the spectral double degeneracy of the Haldane-insulator, which is remarkably maintained at low on-site interaction, where 
triple or higher site occupation is significant and particle-hole symmetry is broken. 
Moreover, surprisingly, the partition dependence of the spectral degeneracy 
in the Haldane-isulator, and of a partial degeneracy in the Mott-insulator, are directly linked to the, in principle unrelated, 
density-density correlations, and presents an intriguing periodic behavior in superfluid and supersolid phases.
\end{abstract}
\date{\today}
\maketitle



Ultra-cold gases in optical lattices~\cite{Maciek-Review,Bloch-Review} allow for the controlled study of 
Hubbard-like models~\cite{Hubbard1963}, fundamental for the description of strongly correlated systems. 
Although until recently only on-site interactions have been relevant, inter-site interactions 
may play a crucial role in new experiments on gases with dipolar interactions~\cite{Thierry-Review}.
Polar lattice gases may allow for the simulation of extended Hubbard models, which play an important role in 
e.g. the analysis of high-$T_c$ superconductivity~\cite{Essler1992}, being the simplest models that capture the interplay between strong 
correlations and charge-ordering effects.

Most remarkably, it has been recently shown that 1D polar lattice bosons, in addition to conventional locally-ordered phases, 
may allow for the so-called Haldane-insulator~(HI) phase~\cite{DallaTorre2006,Berg2008,Dalmonte2011}~(a Haldane-liquid phase has been 
recently proposed for polar fermions~\cite{Kestner2011}). The HI lacks local order, 
but presents non-local string-like order~\cite{Nijs1989}, resembling to a large extent the Haldane phase in integer spin chains~\cite{Mikeska-Review}.
The Haldane phase is one of the paradigms of topological quantum phases, which due to the lack of local order parameter fall beyond the Landau paradigm 
of condensed-matter physics. Topological phases have attracted a growing interest in recent years, most relevantly their entanglement properties. 
In particular, topological phases present an anomalous entanglement entropy, characterized by a topology-dependent universal additive constant~\cite{Kitaev2006,Levin2006}. 

The entanglement entropy is defined as the von Neumann entropy, $S=-Tr\rho_A\log\rho_A$, 
associated to the reduced density matrix, $\rho_A$, obtained after partitioning the 
system into two parts, $A$ and $B$, and tracing out $B$. 
However, $\rho_A$ contains much more information than its mere entanglement entropy. 
In particular, it has been shown that the set of eigenvalues of $\rho_A$, the so-called entanglement spectrum (ES), 
provides key insights concerning quantum correlations in many-body systems~\cite{Li2008}. 
The intriguing properties of the ES in different physical systems have attracted recently 
a major interest~\cite{Calabrese2008,Lauchli2010,Poilblanc2010,Fidkowski2010}.
Interestingly, it has been shown that the ES of the Haldane phase in 1D integer-spin chains 
is characterized by a double degeneracy protected by a set of symmetries~\cite{Pollmann2010}. 
Moreover, the difference between the two largest eigenvalues of the ES~(Schmidt gap) presents at the boundary 
of the Haldane phase in 1D $S=1$ spin chains scaling properties which coincide to those
of the corresponding local order parameter in non-topological phases~\cite{DeChiara2011}.

In this Letter, we consider the 1D extended Bose-Hubbard model~(EBHM) to illustrate 
the interesting, and somewhat unexpected, insights provided by a careful study of the ES.
As for the Haldane phase, the HI phase is characterized by a doubly-degenerated ES. Our numerical results show 
that, remarkably, this double degeneracy~(as suggested in~\cite{Pollmann2010}) is maintained at low on-site repulsions, 
where significant triple, or higher, site occupation breaks particle-hole symmetry. 
We show that, as recently discussed~\cite{Kestner2011,Salerno2010}, 
the analysis of the block structure of $\rho_A$ may offer interesting insights. In particular, 
the block symmetry of $\rho_A$ permits an intuitive understanding of the 
doubly-degenerated ES of the HI, unveiling partial double degeneracy for 
the Mott-insulator~(MI). Moreover, we show that 
the partition-dependence of the ES in finite systems, which to 
the best of our knowledge was not yet considered, provides interesting surprising insights 
on the relation between ES and excitation spectrum. In particular, the Schmidt gap in the HI, and a similar one for 
the MI, present an exponential decay in its partition-dependence, 
with a characteristic length which, remarkably, is identical to that of the, a priori unrelated, 
density-density correlations. Moreover, the partition-dependence 
provides interesting information about other phases, presenting an intriguing periodicity 
in superfluid~(SF) and supersolid~(SS) phases.


We consider below the 1D EBHM  
\begin{eqnarray}
H &=& -t \sum_{i} (b^{\dagger}_i b_{i+1} +  h.c.) \nonumber \\
&+&
 \frac{U}{2} \sum_{i}  n_i(n_i -  1) + V \sum_{i} n_i n_{i+1},
\label{eq_EBHH}
\end{eqnarray}
where $b^{\dagger}_i$, $b_i$ are creation and annihiliation operators for bosons at site $i$, $n_i=b_i^\dag b_i$, 
$t$ is the hopping amplitude, and $U$ and $V$ are, respectively, the coupling constants 
for on-site and inter-site interactions. As mentioned above, inter-site interactions 
appear naturally for polar particles 
in lattices. Although for polar lattice gases interactions beyond 
nearest neighbors may play a role at low energies, 
model~\eqref{eq_EBHH} contains already most of the relevant physics.


The Holstein-Primakoff transformation $S^+_i=(2S-n_i)^{1/2}b_i$, $S^-_i=b^{\dagger}_i(2S-n_i)^{1/2}$ 
and $S^z_i=S-n_i$ links the EBHM with the anisotropic Heisenberg spin-$S$ model:
\begin{eqnarray}
H=\sum_i \left[ J \sum_{j=x,y}S^j_iS^j_{i+1}+J_z S^z_iS^z_{i+1}+D(S^z_i)^2\right]
\label{eq_Heisenberg}
\end{eqnarray}
where $J_z/J=VS/2t$ and $D/J=US/4t$ characterize, respectively, the 
exchange and single-ion anisotropy. 
Assuming $\langle S_z \rangle =0$ (i.e. $\langle n_i \rangle=S$), both models are equivalent 
if $\delta n_i=n_i-S\ll S$ and the maximal site occupation $n_{max} \leq 2S$.  
Below we consider $\langle n_i \rangle =1$ which relates to $S=1$. 
However, since $S=1$ does not fulfill the aforementioned criteria  
significant departures between both models may be expected, especially for low $U/t$.

For $D/J>0$ and $J_z/J>0$, model~\eqref{eq_Heisenberg} presents 
three possible phases~\cite{Mikeska-Review}. 
For sufficiently large $D$, $S_z=0$ is favoured, and the system 
enters the large-D phase, where the N\'eel order  
$O^{\alpha=x,y,z}_{\rm N\acute{e}el}\equiv\lim_{\mathop{|i-j|\to\infty}}(-1)^{|i-j|}\langle S^{\alpha}_i S^{\alpha}_j\rangle=0$, and the non-local string order 
$O^{\alpha}_{\rm string}\equiv -\lim_{|i-j|\to\infty}\langle S^{\alpha}_i e^{i\pi\sum^{j-1}_{i+1}S^{\alpha}_l}S^{\alpha}_j\rangle =0$. 
For sufficiently large $J_z>0$ antiferromagnetic ordering is preferred, and the N\'eel phase occurs, where 
$O^{x,y}_{\rm N\acute{e}el}=O^{x,y}_{\rm string}= 0$ but 
$O^{z}_{\rm N\acute{e}el} \neq 0$ and $O^{z}_{\rm string}\neq 0$.
In between the N\'eel and large-D phases, the Haldane phase is found, a gapped phase 
with antiferromagnetic spin order~($O^{\alpha}_{\rm string}\neq 0$)
but no spatial order~($O^{\alpha}_{\rm N\acute{e}el}=0$). 

At large-enough $U/t$, the EBHM phases for 
$\langle n_i \rangle=1$ map those of $S=1$ chains. 
The equivalent of the large-D phase is a MI 
with one boson per site, with string order parameter~\cite{DallaTorre2006}
$O_{\rm string}\equiv \lim_{r\rightarrow\infty}\langle (1-n_0) 
\textmd{e}^{i\pi\sum^r_{k=0}{n_k}}(1-n_r)\rangle=0$, 
density-density order parameter
$C_{nn}\equiv \lim_{r\rightarrow\infty}[\langle n_0 n_r \rangle - \langle n_0 \rangle \langle n_r \rangle]=0$, 
and parity order parameter~\cite{Berg2008} 
$O_{\rm parity}\equiv \lim_{r\rightarrow\infty}\langle \textmd{e}^{i\pi\sum^r_{k=0} n_k} \rangle\neq 0$.
A density wave (CDW) of alternate sites with $0$ and $2$ bosons is the counterpart of 
the N\'eel phase, with  $O_{\rm string}\neq 0$, $O_{\rm parity}\neq 0$, and $C_{nn}\neq 0$. 
The equivalent of the Haldane phase, the HI~\cite{DallaTorre2006}, 
has $O_{\rm parity}=C_{nn}=0$, $O_{\rm string}\neq 0$.
For low $U/t$, the system may be SF or SS, 
with polynomically decaying single particle correlation.


We investigate the EBHM using density-matrix renormalization group~(DMRG)~\cite{White1992} with up to 
$L=100$ sites, keeping $\sim 300$ optimal states. 
Since the EBHM is number conserving, we implement our DMRG  
for sectors with a specific boson number. 
We consider maximally $4$ bosons per site, which we have checked to be enough for $U/t>0.5$.
A proper investigation of the bulk properties of the HI using open boundary conditions requires avoiding 
the spurious effects introduced by edge states~\cite{Kennedy1990}. We have employed two 
independent ways of handling edge states, ensuring that our results are method independent.
A first method consists in placing $N=L+1$ bosons, 
resulting in a magnetization $\sum_iS^z_i=-1$ which, due to the Haldane gap, 
is accomodated at the edges, $S^z_1=S^z_{L}=-1/2$, with no energy cost.  
The edge states are then polarized.
An alternative method consists in adding a site with a hard-core boson at each chain boundary 
(sites $0$ and $L+1$). This method 
is equivalent to that employed in $S=1$ in Ref.~\cite{White1993b}. By properly 
tuning the coupling between the bulk and the extra sites, we may achieve 
$\langle n_{0}\rangle=\langle n_{L+1}\rangle \approx 1/2$, i.e. the same  
probability of effective spin-$1/2$ up and down states. The spin-$1/2$ forms a singlet 
with the edge spins, destroying the edge states. Note, however, that the incommesurability 
associated with the handling of the edge states leads to a commesurate-incommesurate 
transition of the MI into SF. The analysis of the MI phase demands hence $L=N$.

\begin{figure}
\resizebox{2.3in}{!}{\includegraphics{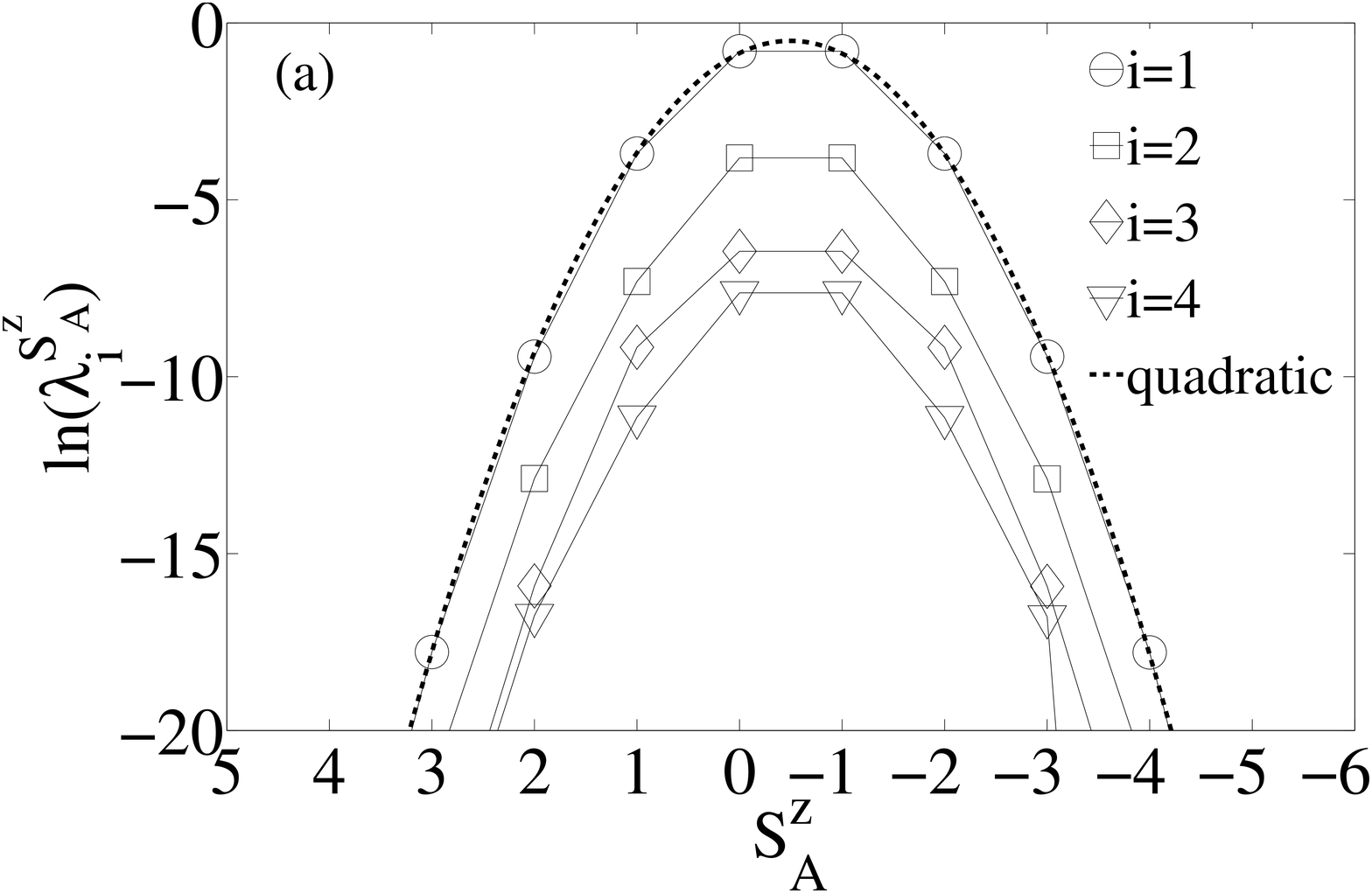}}
\resizebox{2.3in}{!}{\includegraphics{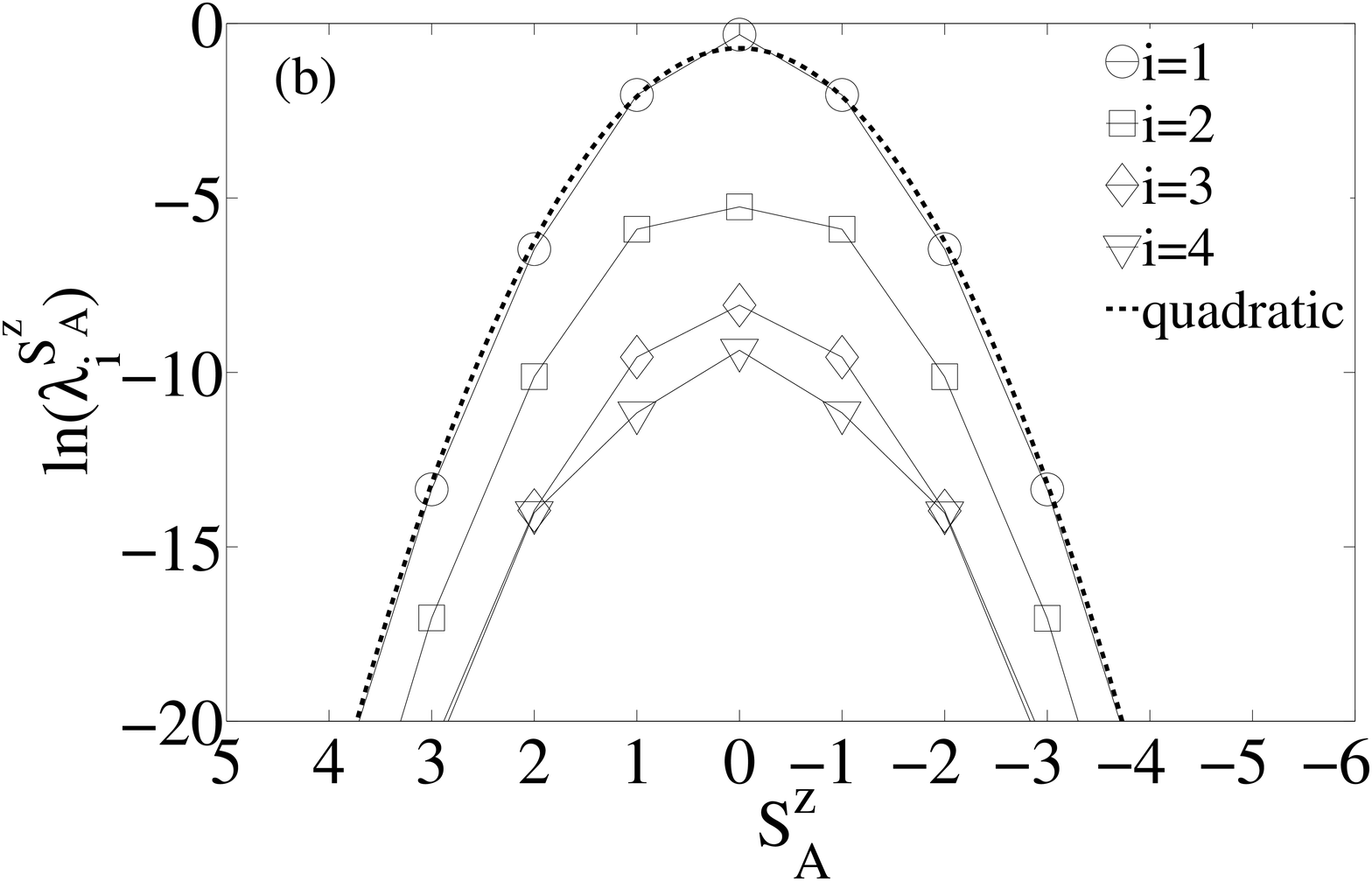}}
\resizebox{2.3in}{!}{\includegraphics{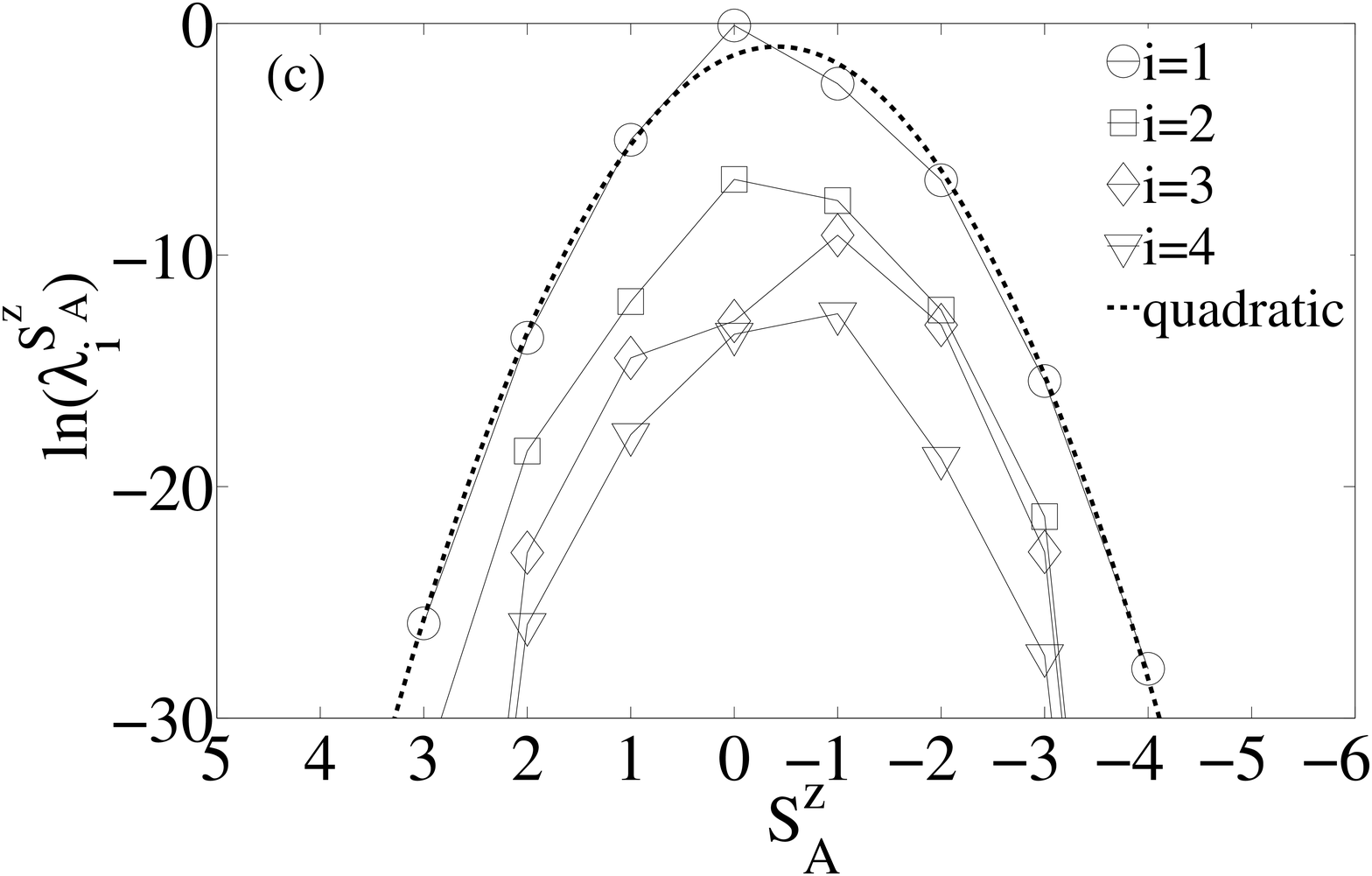}}
\caption{Largest eigenvalues of the ES $\lambda_i^{S^z_A}$
for different $S^z_A$ sectors, for $U/t=5$, $L=100$, 
and $V/t=3.3$ (HI) (a), $2.0$ (MI) (b) and $4.0$ (CDW) (c). The results for $\lambda_1^{S^z_A}$ 
are compared with a Gaussian distribution (dotted).}
\label{fig:1}
\end{figure}

We recall that the ES is obtained after dividing the system into two parts, $A$ (with $L_A$ sites) and $B$, tracing out $B$, 
and diagonalizing the reduced density matrix $\rho_A$. 
Note that due to number conservation in the whole system, 
$\rho_A$ is block diagonal, with blocks $\rho (S_A^z)$ characterized by an effective magnetization $S_A^z=L_A-N_A$ 
given by the number of particles $N_A$ in $A$. The ES is hence grouped in sectors with a given $S_A^z$, 
$\lambda_{i}(S^z_A)$, with $i$ the index of the eigenvalue~(in decreasing order). 
The structure of $\lambda_{i}(S^z_A)$~(Figs.~\ref{fig:1}) 
offers an interesting insight into the degeneracies of the ES.  
Fig.~\ref{fig:1}(a) shows that, as for the Haldane phase of the 
$S=1$ chain~\cite{Pollmann2010}, a doubly degenerate ES is recovered for the HI phase. 
Fig.~\ref{fig:1}(a) shows that the double degeneracy 
is directly linked to the symmetry $\rho(S_A^z+s)=\rho(S_A^z-s-1)$, with $s=0,1,\dots$~\cite{footnote1}, 
which may be linked to number conservation and bond-centered symmetry~\cite{Pollmann2010}. 
Remarkably, we observe a basically perfect double degeneracy in the HI phases even at $U/t$ as low as $1$, where 
occupations $n_i>2$ are very significant, hence clearly breaking particle/hole symmetry.
Finally note that $\{\lambda_1(S_A^z)\}$ present a Gaussian dependence around $S^z=-1/2$, 
resembling the ES of permutationally-symmetric models~\cite{Salerno2010}, and that 
blocks with $S_A^z>1$($<-2$) appear due to solitons in the string-order~\cite{Yamamoto1997} and sites with $n_i>2$.

Interestingly, the ES of the MI~(Fig.~\ref{fig:1}(b)) 
presents block symmetry as well, $\rho(S_A^z+s)=\rho(S_A^z-s)$. 
However, the eigenvalues of $S_A^z=0$ are not degenerated. 
As a result, the spectrum of the MI presents a mixture between 
a partial double degeneracy and non-degenerate eigenvalues. In particular, a gap is observed between 
the two largest eigenvalues of the spectrum (Schmidt gap~\cite{DeChiara2011}), whereas the 
second and third eigenvalues are degenerated. 
Finally, in the CDW the block symmetry is lost~(Fig.~\ref{fig:1}~(c)), and 
hence the ES does not present any systematic degeneracy. 
All cases show an approximate Gaussian dependence of $\lambda_1^{S_A^z}$, which may 
be employed for designing more efficient DMRG codes.


\begin{figure}
\resizebox{2.1in}{!}{\includegraphics{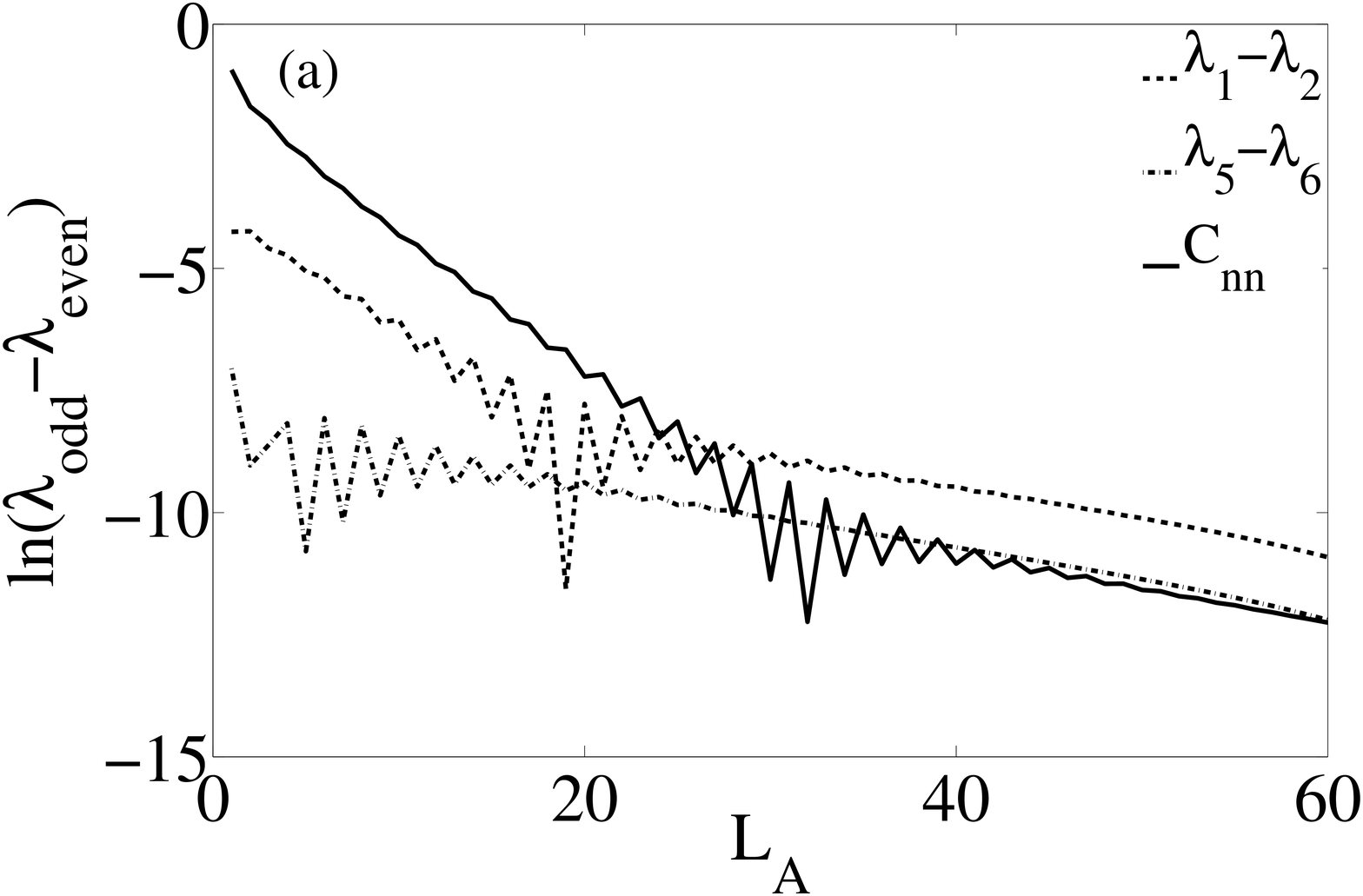}}
\resizebox{2.1in}{!}{\includegraphics{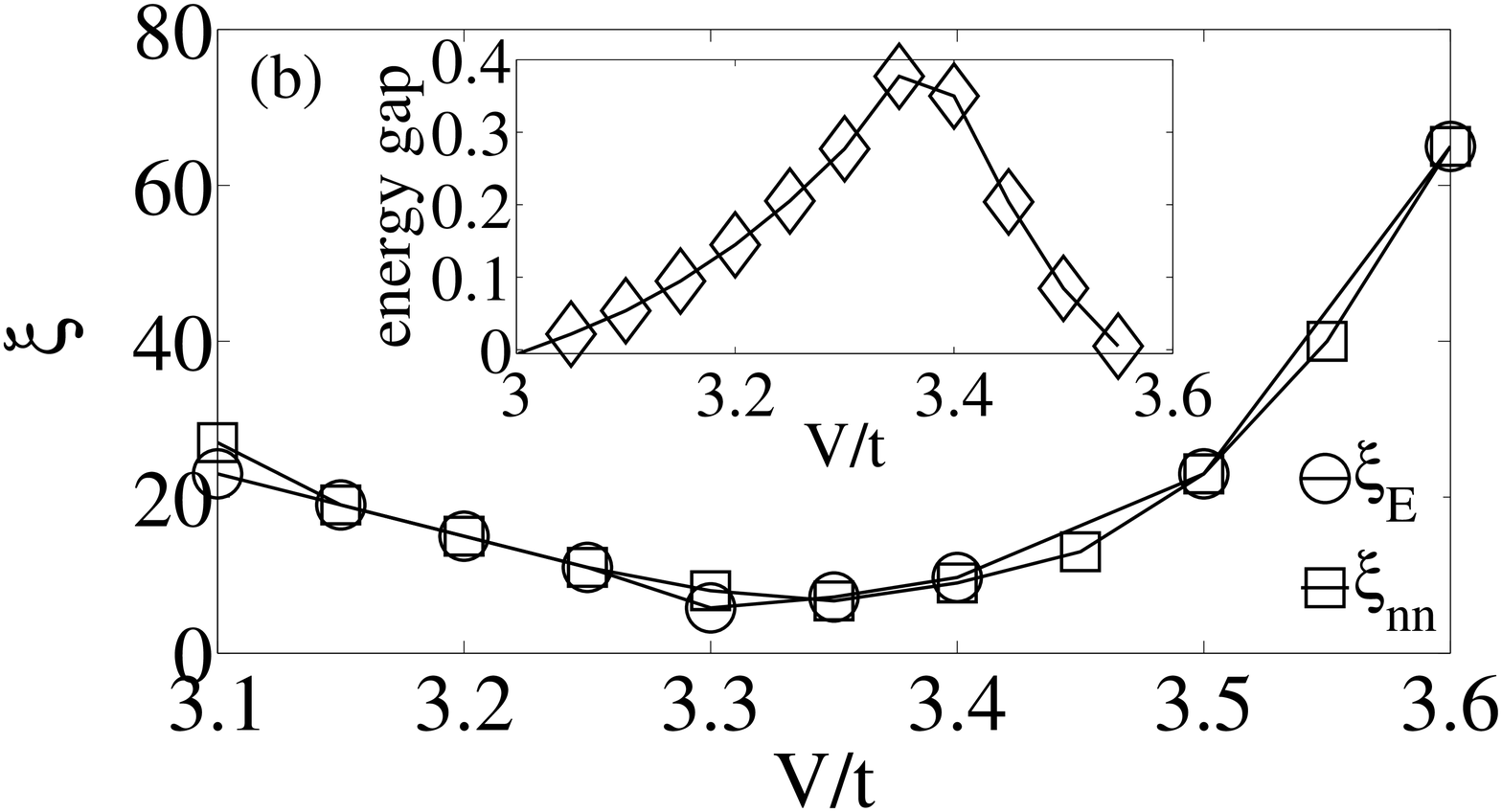}}
\resizebox{2.1in}{!}{\includegraphics{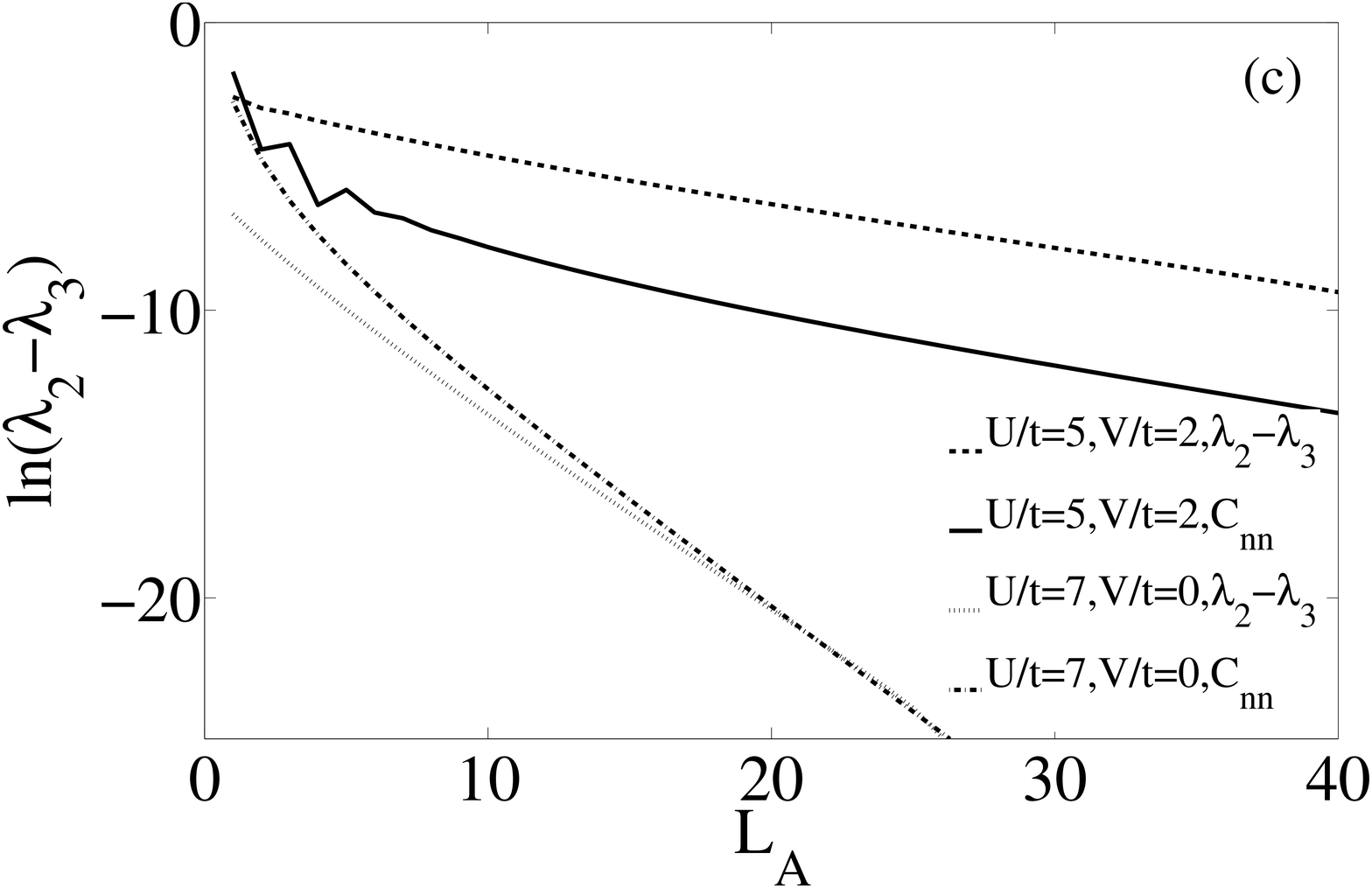}}
\caption{ (a) Partition dependence of the Schmidt gap~(and $\lambda_5-\lambda_6$) 
for $L=150$, $U/t=5$, and $V/t=3.2$ (HI), compared to the 
density-density correlation $C_{nn}(i)$; (b)
Decay length of the Schmidt gap~($\xi_E$) and $C_{nn}(i)$~($\xi_{nn}$) 
as a function of $V/t$ inside the HI phase, (inset) Haldane-like gap for number-conserving excitations 
after extrapolation to $L\rightarrow\infty$; 
(c) Partition dependence of $\lambda_2-\lambda_3$ compared to $C_{nn}(i)$ for $L=N=100$, for 
two MI cases.}
\label{fig:2}
\end{figure}


In finite systems, the ES depends on the position of the partition, i.e. on $L_A$. 
Interestingly, as we show in the following, this $L_A$ dependence  
provides interesting non-trivial information about both locally and non-locally ordered phases.
We plot in Fig.~\ref{fig:2}(a) the Schmidt gap $\Lambda(L_A)\equiv\lambda_1-\lambda_2$ 
between the two largest eigenvalues of $\rho_A$ for different values of $L_A$, 
starting from $L_A=1$. 
Sufficiently away from the chain borders, we find that in the HI $\Lambda (L_A)$ and also all 
$\lambda_{2j-1}-\lambda_{2j}$ decay exponentially. Remarkably, as shown in Fig.~\ref{fig:2}(b), 
the typical length characterizing this exponential decay, $\xi_E$, 
is identical to the correlation length, $\xi_{nn}$, associated to the exponentially decaying density-density 
correlation in the bulk $C_{nn}(i)\equiv\langle n_{j}n_{j+i} \rangle -\langle n_j \rangle\langle n_{j+i}\rangle$. 
Since $\xi_{nn}$ is inversely proportional to the Haldane-like gap for number-conserving excitations~(inset in Fig.~\ref{fig:2}(b)), 
our results show that there is a non-trivial, and somewhat unexpected, link between 
the partition dependence of the ES and the energy gap characterizing the 
excitation spectrum. 
Moreover, a similar analysis for the $L_A$ dependence of the ES
may be performed for the gap $\tilde\Lambda(L_A)\equiv\lambda_2- \lambda_3$ 
in the MI phase. Our analysis shows that $\tilde\Lambda(L_A)$ also decays exponentially, 
with a characteristic length identical to that characterizing the exponential 
decay of $C_{nn}(i)$~(Fig.~\ref{fig:2}(c)), 
and hence inversely proportional to the Mott gap. Note that the relation between $L_A$ dependence 
of ES degeneracies and density-density correlations is also particularly interesting since it may allow to evaluate 
correlation lengths with a minimal effort in DMRG simulations.

\begin{figure}
\resizebox{2.2in}{!}{\includegraphics{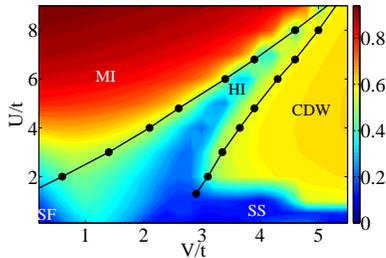}}
\caption{Density plot of $\zeta$ (see text) in the plane $U/t$, $V/t$, for $L=N=61$. 
In the diagram we identify different phases, including MI, 
HI, CDW, SF and SS. The circles indicate the phase boundaries in a $S=1$ spin chain.}
\label{fig:3}
\end{figure}
\begin{figure}\resizebox{2.2in}{!}{\includegraphics{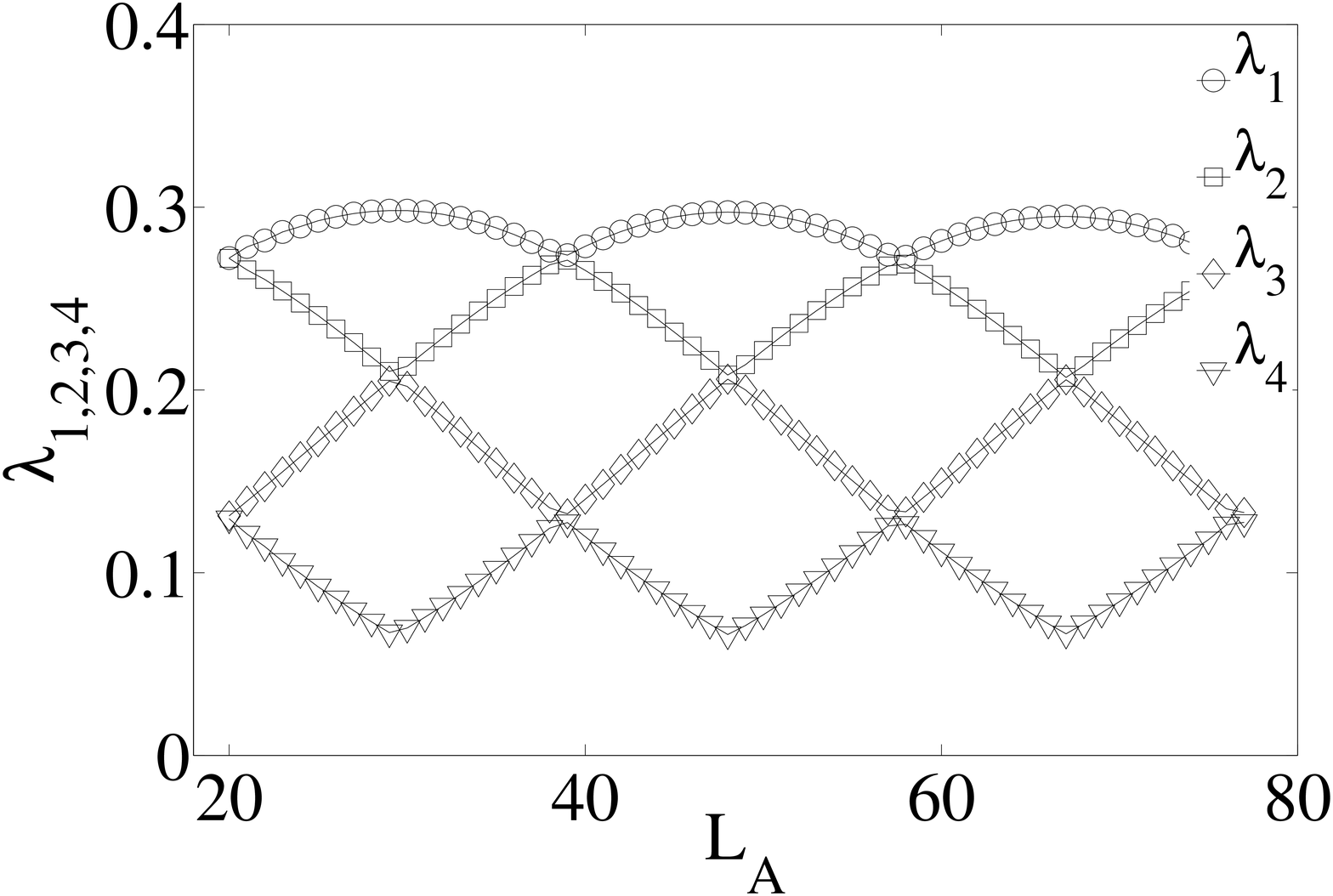}}
\caption{Entanglement spectrum as a function of the partition $L_A$ for $L=100$, $U/t=0.5$ and $V/t=3.$.}
\label{fig:4}\end{figure}

By means of the ES we may characterize the phase diagram of the EBHM as a function of $U/t$ and $V/t$. 
We evaluate the largest eigenvalues $\lambda_j(L_A)$ of $\rho_A$ for different $L_A$ values, and 
define $\lambda_j^T\equiv \frac{1}{L}\sum_{L_A=1}^L \lambda_j(L_A)$. 
We then employ the four largest eigenvalues to define 
$\zeta\equiv \lambda^{T}_1-\lambda^{T}_2+\lambda^{T}_3-\lambda^{T}_4$, 
which characterizes the degeneracy of the entanglement, being zero for 
perfect double-degeneracy of the ES. Proceeding in this way for different values of $U/t$ and $V/t$ we obtain the results depicted in 
Fig.~\ref{fig:3}. Note that the figure shows clearly the appearance of different regions, although a simultaneous analysis of MI and HI 
demands the use of $N=L$ particles, which prevents the destruction or polarization of the edge states, 
and as a result the ES in the HI region is not fully degenerated. For $U/t>2$ the boundaries are in very good agreement 
with our results obtained from correlation functions.
In Fig.~\ref{fig:3} we depicted as well the results expected for a $S=1$ chain for the boundaries between Haldane, large-D and N\'eel phases, 
which are in good agreement with the results for $\zeta$ for $U/t>2$. For $U/t<2$ it is difficult to get a reliable 
information about the decay of correlations. However, the ES provides interesting information about the system even at low $U$. 
For small $U/t$ a clear separate region (the SF region) occurs for small $V/t$. 
Moreover, we find that at low $U$ the region of low $\zeta$ significantly extends 
inside the CDW region. We identify this region with the apperance 
of a supersolid phase~\cite{Mishra2009}. Interestingly, this region presents a peculiar $L_A$ dependence 
of the ES~(Fig.~\ref{fig:4}) characterized by a a periodic modulation with a progressively shorter period 
for growing $V/t$ when approaching the CDW phase.  
We have found these intriguing dependence also for SF phases at low $V/t$ with non-commesurate fillings.



In summary, the block structure of the ES and its partition dependence  
in finite systems provide interesting, and somewhat unexpected, insights about correlated quantum systems.
This has been illustrated for a 1D EBHM, for which we have shown 
that block symmetry provides an intuitive understanding about spectral degeneracy in the HI, 
which remarkably persists when low on-site repulsion breaks particle-hole symmetry.
Our results show as well that the partition-dependence of the ES in finite systems is surprisingly linked 
to density-density correlations in both MI and HI, providing as well interesting insights 
about other phases. Finally, we note that our findings concerning block symmetry and partition-dependence are expected 
to be applicable to other models, for both locally- and non-locally-ordered phases.

We thank T. Vekua and E. Orignac for discussions, and J. Kestner for calling our attention about Ref.~\cite{Kestner2011}.
We acknowledge the support from the Centre for Quantum Engineering 
and Space-Time Research QUEST.

\end{document}